\documentclass[prb,twocolumn,showpacs,amsmath,superscriptaddress,aps]{revtex4}
\usepackage{graphicx}
\graphicspath{{./figures/}}
\DeclareGraphicsExtensions{{.eps}}
\usepackage{dcolumn}
\usepackage{bm}
\begin{document}

\title{Electronic transport through domain walls in ferromagnetic nanowires:\\
       Co-existence of adiabatic and non-adiabatic spin dynamics}

\author{Victor A. Gopar}
\affiliation{Institut de Physique et Chimie des Mat\'{e}riaux de Strasbourg, 
             UMR 7504 (CNRS-ULP), 23 rue du Loess, 
             BP 43, 67034 Strasbourg Cedex 2, France}
\affiliation{Institut f\"{u}r Theorie der Kondensierten Materie, 
             Universit\"{a}t Karlsruhe, Postfach 6980, 
             76128 Karlsruhe, Germany}
\author{Dietmar Weinmann}
\affiliation{Institut de Physique et Chimie des Mat\'{e}riaux de Strasbourg, 
             UMR 7504 (CNRS-ULP), 23 rue du Loess, 
             BP 43, 67034 Strasbourg Cedex 2, France}
\author{Rodolfo A. Jalabert}
\affiliation{Institut de Physique et Chimie des Mat\'{e}riaux de Strasbourg, 
             UMR 7504 (CNRS-ULP), 23 rue du Loess, 
             BP 43, 67034 Strasbourg Cedex 2, France}
\author{Robert L. Stamps}
\affiliation{Department of Physics, University of Western Australia, 
         Nedlands WA 6907, Australia}

\begin{abstract}
We study the effect of a domain wall on the electronic transport in 
ferromagnetic quantum wires. Due to the transverse confinement,
conduction channels arise. In the presence of a domain wall, spin up
and spin down electrons in these channels become coupled.
For very short domain walls or at high longitudinal 
kinetic energy, this coupling is weak, leads to very few spin flips, 
and a perturbative treatment is possible. 
For very long domain wall structures, the spin follows adiabatically
the local magnetization orientation, suppressing the effect of the
domain wall on the total transmission, but reversing the spin of the
electrons. In the intermediate regime, we numerically investigate the 
spin-dependent transport behavior for different shapes of the domain wall. 
We find that the knowledge of the precise shape of the domain wall 
is not crucial for determining the qualitative behavior.
For parameters appropriate for experiments, electrons with low
longitudinal energy are transmitted adiabatically while the electrons
at high longitudinal energy are essentially unaffected by the domain
wall. Taking this co-existence of different regimes into account
is important for the understanding of recent experiments. 
\end{abstract}


\maketitle

\section{Introduction}

A new kind of electronic devices taking advantage of the electron 
spin have been developed during the last years. 
The influence of the spin on electronic transport attracts considerable
interest since early experiments in multi-layered magnetic structures 
have shown that the resistance is considerably increased in the case 
of an anti-parallel magnetization of the layers, as compared to a 
parallel configuration \cite{baibich,binasch}. This is at the base of the 
so-called Giant Magneto-Resistance (GMR), which is already used in
the read-heads of commercial high performance hard-disks. 

In magnetic configurations that are obtained when one
substitutes the non-magnetic spacer layer between the ferromagnets by
domain walls \cite{gregg}, the effect of a magnetic domain wall on 
the electronic transport properties has become a subject of 
great interest. In particular, the effect of a {\it single domain wall} 
on the resistance of a ferromagnetic nanowire has been measured
for electro-deposited cylindrical Co wires down to 35 nm in 
diameter \cite{ebels}, and thin polycrystalline Co films having 
a thickness of 42 nm and a width down to 150 nm \cite{dumpich}. 
The results of both experiments indicate that, besides negative
contributions from the anisotropic magneto-resistance, the domain wall 
scattering yields a positive contribution to the resistance.

The prospect of interesting technological applications 
of magneto-electronic devices exploiting the spin degree of freedom of
the electrons together with its importance from the fundamental
point of view has strongly motivated theoretical studies of 
spin-dependent electronic transport.

Many efforts have been made in order to explain the electronic 
transport, in particular the enhancement of the magneto-resistance 
in these spin-dependent electronic devices. 
For example, a Boltzmann equation has been applied 
to study the resistance of multi-layered magnetic/non-magnetic
structures when the spin-diffusion length is larger than the mean 
free path \cite{valet}.
While the electronic spin is expected to follow adiabatically a 
very slowly varying magnetization \cite{stern}, the deviations from
this adiabatic behavior, which are due to the
finite length of the domain wall, lead to a so-called mistracking of
the spin, and result in a GMR-like enhancement of the 
magneto-resistance in multi-domain wall configurations \cite{gregg}. 
Including spin-dependent scattering, the mistracking and the
resulting magneto-resistance have been calculated for such a
system, within the so-called two-band model \cite{levy}, which 
consists in a simplification of the complicated band structure
of a ferromagnetic metal. 

An outstanding problem in magnetism is a fully consistent description 
of transport and thermal properties in terms of electronic states 
calculated from first principles. Despite remarkable progress in the
past twenty years, such a description does not yet exist in a form
suitable for predicting features such as domain wall structures in 
non-equilibrium situations. As such, it is reasonable to search for
suitably simplified model descriptions that capture the essence of the
important physics involved. In the present case, a two-band model is
useful for the study of how the geometrical characteristics of a
magnetic domain wall affect electron transport. We make a distinction 
between spatially extended electronic states that contribute strongly
to conduction and more localized states that contribute strongly to
the formation of local magnetic moments. For the transition metals,
this model assumes that exchange correlations between electrons in 
primarily $d$-like orbitals are largely responsible for the formation 
of magnetic moments leading to the microscopic magnetization. The
$s$-like orbitals contribute much less strongly to the magnetization 
and instead interact relatively weakly with the local moments via a contact
interaction term. These types of $s$-$d$ interaction models have proven
very useful in the past for discussions of indirect exchange
interactions in magnetic transition metal multilayers.

In our work we therefore assume that the domain wall represents a
stable magnetic state with an energy above the ferromagnet ground
state. The exact shape and dimensions of the wall are determined
by exchange correlation energies and spin orbit interactions primarily 
affecting electrons associated with the magnetization of the wall. 
These interactions are small perturbations on the conduction electron
states, and the interaction between conduction electrons and the
domain wall is represented by a simple contact potential. In a single 
electron picture, the wall appears as a spatially varying spin
dependent potential for the conduction electrons. The magnitude
of the splitting between the spin up and spin down potentials is taken 
as a free parameter that is related to the exchange correlation energy 
of the electrons involved in forming the wall, and the contact
potential describing interaction of conduction electrons with the
effective potential associated with the wall structure. In the
following we refer to this contact potential between conduction
electrons and the magnetization as an 'exchange interaction' although 
it is quite distinct from the exchange interaction used to
parameterize the interactions leading to magnetic ordering. 

In thin ballistic quantum wires and narrow constrictions or point
contacts, the lateral confinement of the electronic wave-functions 
leads to the emergence of quantized transport channels. As a 
consequence, the conductance is quantized and exhibits steps of
$e^2/h$ as a function of the Fermi energy \cite{wees,wharam}.   
Nakanishi and Nakamura \cite{nakanishi} considered the
conductance of very narrow quantum wires including the effect of 
a domain wall. A perturbative approach allowed them to study the
effect of a very short domain wall on the conductance steps.
Imamura and collaborators \cite{imamura} calculated the conductance 
of a point contact connecting two regions of a ferromagnet having 
parallel or anti-parallel magnetization directions. Within an $s-d$
two-band model, they numerically obtained a non-monotonic dependence 
of the domain wall contribution to the resistance on the width of 
the point contact. 

There have been attempts to compare different approaches to the
calculation of the domain wall magneto-resistance (DWMR), which point
towards the importance of including more realistic band structures. 
In the ballistic case, van Hoof and collaborators \cite{hoof}
calculated the DWMR for an adiabatic model where the magnetization 
direction changes very slowly along the wire, using an extension 
of the standard band structure calculation to include an infinite 
spin spiral, as well as a ``linear'' model, where the magnetization turns 
in a finite region at a constant rate. These two approaches yield a
much larger effect than a two-band model, where corrections with 
respect to an infinitely long domain wall are calculated. On the other 
hand, first principle calculations for the case of abrupt 
magnetization interfaces yield a DWMR which is orders of magnitude 
larger than for the other models that take into account realistic 
domain wall lengths. 

It then seems necessary to develop more accurate treatments within the
two-band model in order to understand the crossover from the abrupt
domain wall situation to the adiabatic regime. Since we are working
with nanowires, obvious transverse quantization effects appear, which 
are more easily tractable within a two-band model. Moreover, the
two-band model allows to easily obtain the transmission coefficients 
with and without spin flip, making it possible to study the
mistracking effect in finite length domain walls and its GMR-like 
consequences. Finally, given the typical experimental parameters, 
it would be important to go beyond the ballistic limit. Taking into 
account disorder within a two-band model seems much more doable than 
in the framework of a band structure calculation.   

In this paper, we study the effect of a domain wall on the 
conductance of a nanowire within the two-band model, comparing
different shapes and sizes of the domain wall. Focusing on the
contribution of the spin-dependent scattering of the domain wall, we
do not consider material-dependent contributions to the
resistance like the anisotropic magneto-resistance.
 
After presenting our model in section \ref{sec:model},
we consider the perturbative regime of weak spin coupling
induced by the domain wall in section
\ref{sec:weak_coupling}. From a comparison of different domain wall
shapes, we shall extract the relevant parameters governing the
transmission through the domain wall with and without spin-flip processes. 
While this applies to short domain walls, in section 
\ref{sec:strong_coupling} we study the transport in the general case 
of a strong spin coupling induced by the domain wall, 
which allows to treat domain walls of arbitrary length.
We present the deviations 
from the full transmission with spin rotation 
following the local magnetic structure expected for infinitely long
domain walls, which is due to finite
domain wall length, and treat the most interesting crossover regime. 
This case is relevant since typical 
experiments \cite{ebels,dumpich} are far from the thin wall regime,
but not really in the adiabatic limit in which the length of 
the domain walls is large. 

\section{Model}\label{sec:model}

\begin{figure}
\begin{center}
\includegraphics[width=\columnwidth]{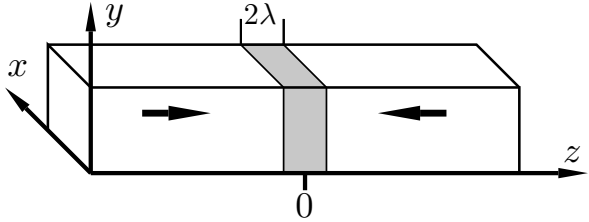}
\caption{\label{fig:wallsketch}Sketch of a
  ferromagnetic quantum wire containing a domain wall (grey region),
  for the example geometry of a square cross-section. The arrows
  indicate the magnetization directions far from the domain wall, for
  the case of a N\'{e}el wall.}
\end{center}
\end{figure}

We consider a wire along the $z$ axis with a domain
wall, and choose the origin ($z=0$) in the middle of the wall 
(see Fig.~\ref{fig:wallsketch}). 
As explained above, we work within the two-band model, where the
$d$ electrons are responsible for the magnetization and the current is
carried by the $s$ electrons. Therefore, we write for the latter an effective
Hamiltonian
\begin{equation}
\label{eq:H}
H=-\frac{\hbar^2}{2m}\nabla^2
   +\frac{\Delta}{2} \vec{f}(\vec{r})\cdot\vec{\sigma}\, , 
\end{equation}
where $\Delta$ is the spin splitting of the $s$ electrons due to the
exchange coupling with the $d$ electrons and $\vec{\sigma}$ is the
vector of the Pauli matrices. 
The unit vector $\vec{f}$ represents the direction of the local 
magnetization. Its functional dependence describes the shape of 
the domain wall. The lateral confinement present in a nanowire may
have a considerable influence on this shape, leading to domain walls
which are altered as compared to the case of bulk domain walls 
\cite{dumpich,bruno,prejbeanu1,prejbeanu2,hausmanns}. 
In addition, a spin-polarized current through the domain wall creates
a torque which can alter its shape \cite{waintal}. Working in the
linear response regime of low current, we do not need to take into
account this back-action of the conduction electrons on the magnetic 
structure.

Assuming that the magnetization only depends on $z$, we will therefore
be interested in comparing different functional forms of the kind  
$\vec{f}=\{f_x(z), 0, f_z(z)\}$. This choice does not imply a 
loss of generality and corresponds to a so-called
``N\'{e}el wall'', where the magnetization is parallel to the wire
axis in the leads far from the domain wall 
(arrows in Fig.~\ref{fig:wallsketch}), and turns inside the $x$--$z$ plane
parallel to this wire axis when going through the wall. 

The assumption $\vec{f}(\vec{r})=\vec{f}(z)$ allows us to separate the 
transverse and longitudinal parts of the Hamiltonian (\ref{eq:H}). 
The transverse quantization gives rise to transport channels with
quantum numbers $n_1$ and $n_2$, and an energy $E_{n_1,n_2}$. The 
density of channels $\rho=2\pi m A/\hbar^2$ (where $A$ is the
cross-section of the transverse area of the wire) is equal to that of 
a two-dimensional system, and thus independent of the energy and the 
shape of the cross-section. For the example of a wire with a square 
cross-section of side $w$, we have
\begin{equation}
E_{n_x,n_y}=\frac{\hbar^2}{2m}\left(\left(\frac{\pi
  n_x}{w}\right)^2+\left(\frac{\pi n_y}{w}\right)^2\right)\, .
\end{equation}   

Far away from the domain wall, the orbital parts of the eigenstates
are products of transverse channels and longitudinal plane
waves. Since $\lim_{z \to \pm\infty}f_z(z)= \pm 1$, 
the associated eigenenergies
for spin up are 
\begin{equation}
E_\uparrow = 
E_{n_1,n_2}+\frac{\hbar^2 k_z^2}{2m}\pm \frac{\Delta}{2}\, , 
\end{equation}
while for spin down we have  
\begin{equation}
E_\downarrow = 
E_{n_1,n_2}+\frac{\hbar^2 k_z^2}{2m}\mp \frac{\Delta}{2}\, .
\end{equation}
The domain wall leads to the scattering of these states and the
conductance (in units of $e^2/h$) is given by the Landauer formula  
\begin{equation}\label{eq:landauer}
g=\sum_{n_1,n_2} \sum_{\sigma,\sigma'}
T_{n_1,n_2}^{\sigma,\sigma'}(E_{\textrm{F}})\, ,
\end{equation}
where the sum is done over the occupied channels.
$T_{n_1,n_2}^{\sigma,\sigma'}$ is the transmission coefficient in the
channel $(n_1,n_2)$, for scattering of electrons with spin $\sigma$
into spin $\sigma'$. Such a coefficient only depends on the
longitudinal energy 
\begin{equation}
\epsilon=E_{\textrm{F}}-E_{n_1,n_2}
\end{equation} 
as
\begin{equation}
T_{n_1,n_2}^{\sigma,\sigma'}(E_{\textrm{F}})=T^{\sigma,\sigma'}(\epsilon)\, .
\end{equation}
Therefore, for each channel $(n_1,n_2)$, we are left with an effective
one-dimensional problem at energy $\epsilon$. 

In order to determine the transmission probability 
$T^{\sigma,\sigma'}(\epsilon)$, we write the spinor wave-function 
in the up-down basis (with fixed, $z$-independent spin orientations) as 
\begin{equation}
|\psi(z)\rangle=
\phi_\uparrow(z)|z,\uparrow\rangle+\phi_\downarrow(z)|z,\downarrow\rangle \, ,
\end{equation} 
where $|z,\uparrow\rangle$ has to be interpreted as the tensor product
of the position eigenvector $|z\rangle$ and the spin up state 
$|\uparrow\rangle$.
The Schr\"odinger equation with the Hamiltonian (\ref{eq:H})
leads to a system of coupled differential equations for the components
$\phi_\uparrow(z)$ and $\phi_\downarrow(z)$:
\begin{subequations}
\label{eq:cde}
\begin{equation}
\label{eq:df1}
\frac{d^2}{dz^2}\phi_\uparrow+
\frac{2m}{\hbar^2}\left(\epsilon-\frac{\Delta}{2} f_z
\right)\phi_\uparrow =\frac{2m}{\hbar^2}\frac{\Delta}{2} f_x \phi_\downarrow
\end{equation}
\begin{equation}
\label{eq:df2}
\frac{d^2}{dz^2}\phi_\downarrow+
\frac{2m}{\hbar^2}\left(\epsilon+\frac{\Delta}{2} f_z
\right)\phi_\downarrow =\frac{2m}{\hbar^2}\frac{\Delta}{2} f_x
\phi_\uparrow \, . 
\end{equation}
\end{subequations} 
While the term containing $f_z$ plays the role of a spin-dependent
potential, the transverse component $f_x$ of the wall profile is 
responsible for the coupling between the spinor components
$\phi_\uparrow$ and $\phi_\downarrow$. The scattering solutions of 
Eq.~(\ref{eq:cde}) are then needed to calculate the transmission
coefficients, and therewith the conductance through the domain wall.

For the extreme case of an abrupt domain wall, when $f_z$ has a jump from $-1$
to $1$, the right-hand-side of Eqs.\ (\ref{eq:cde}) vanishes, and spin
up and spin down electrons remain uncoupled. The only effect of the 
discontinuity of $f_z$ is a spin-dependent potential step of height 
$\pm\Delta$ for spin up/down electrons. 
Incoming spin up electrons having longitudinal energy $\epsilon<\Delta/2$ 
cannot overcome this step and are reflected with probability one.
Since the density of conduction channels is independent of the energy 
for wires having a two-dimensional cross-section, this mechanism blocks a
fraction $\Delta/2E_{\rm F}$ of the conduction channels
\cite{wsj_moriond}, all of which exhibit perfect transmission in the
absence of the domain wall.
If one neglects the effect of the potential step
on electrons having higher longitudinal energy ($\epsilon>\Delta/2$), 
this channel blocking mechanism leads to a relative change in conductance
\begin{equation}
\frac{\delta g}{g}=-\frac{\Delta}{2E_{\rm F}}\, ,
\end{equation}  
due to the presence of the domain wall. Taking into account the spin 
conserving reflections for $\epsilon > \Delta/2$ leads \cite{falloon} 
(in the limit $E_{\rm F}\gg \Delta$) to an increase of the effect by a 
factor $4/3$.

The precise shape of the actual domain wall present in an experimental
measurement (which is very difficult to know) would in principle be 
needed to determine the scattering states.
In addition, it is not possible to find an analytical solution of
Eqs.\ (\ref{eq:cde}) for arbitrary domain walls. This is why we introduce
various models of a domain wall, and approximate analytical, as well
as numerical calculations. 
 
In the bulk, when the magnetization always remains parallel to the
domain wall, we have the so-called Bloch walls, whose shape was 
originally calculated by minimizing the total free energy in the 
thermodynamic limit \cite{landau}. In this case, 
$\vec{f}(\vec{r})$ is given by 
\begin{equation}\label{eq:bloch_wall}
\vec{f}(z)=\left\{ \tanh\left(\frac{z}{\lambda}\right), 
\mathrm{sech}\left(\frac{z}{\lambda}\right), 0\right\} \, ,
\end{equation}
where $\lambda$ is the length scale of the domain wall. The lateral
confinement present in a nanowire will certainly alter the previous
functional form of $\vec{f}$. Moreover, an easy magnetization axis in
the direction of the nanowire will result in a N\'{e}el wall, modified
by the transverse confinement. Such effects have been recently
discussed in the literature 
\cite{prejbeanu1,prejbeanu2,hausmanns}.

The variety of possible domain wall structures motivates us to
consider different domain wall profiles: linear, trigonometric and 
extended (defined below), in order to determine the influence of the 
domain wall shape on the conductance of the wire. As we will see
below, while most of the effects are not very sensitive to the 
details of the wall, the signature of the particular domain wall 
appears in some regimes. 

A possible starting point is to assume that the N\'eel-like
confined domain wall has components with the same functional form as
in Eq.\ (\ref{eq:bloch_wall}). In this case we will consider the
``extended'' domain wall defined by the magnetization direction
\begin{equation}\label{eq:extended_wall}
\vec{f}^{(\mathrm{ex})}(z)=
\left\{ {\rm sech}\left(\frac{z}{\lambda}\right), 0 , 
\tanh\left(\frac{z}{\lambda}\right) \right\} \, .
\end{equation}
As compared with the situation of a Bloch wall, this leads to a
permutation of the spatial variables in Eqs.\ (\ref{eq:cde}), and does
not change the results for the transmission coefficients. This is why
in the limit of high electron energy we can compare our results with
the ones of Cabrera and Falicov \cite{cabrera}, who considered Bloch 
domain walls.

In the case of weak coupling between the spin up and down states
described by Eq.\ (\ref{eq:cde}) and short domain walls it is
reasonable to approximate $f_z$ in the wall profile with a linear 
function of position
\begin{equation}\label{eq:linear_wall}
\vec f^{(\mathrm{lin})}(z)=\left\{
\begin{array}{cl}
\left\{\sqrt{1-(z/\lambda)^2},\;0,\; z/\lambda \right\},&
{\rm for}\; |z|< \lambda \\
\left\{\;0 \; ,\;\; 0\;\; , \;{\rm sgn}(z)\; \right\}, &{\rm for}\; 
|z| \ge \lambda .
\end{array}\right.
\end{equation}
The semi-circle form of the coupling term $f_x$ in this ``linear''
wall ensures that $|\vec{f}(z)|^2=1$. This is an important difference 
as compared to our previous work \cite{wsj_moriond}, where the 
conservation of the absolute value of the magnetization was not respected. 
The above constraint has only quantitative consequences in the short 
wall limit which play a role when comparing different wall profiles, 
but becomes crucial for longer domain walls in the adiabatic regime. 

For an arbitrary domain wall, the extension of
the standard recursive Green function method \cite{lee,pastawski}  
to take into account the spin degree of freedom (in a tight binding setup)
allows us to calculate the transmission and 
reflection coefficients $T_{\uparrow\uparrow}$, $T_{\uparrow\downarrow}$,
$R_{\uparrow\uparrow}$ and $R_{\uparrow\downarrow}$.

The case of a ``trigonometric'' domain wall
\begin{equation}
\label{eq:trigonometric_wall}
\vec f^{(\mathrm{tri})}(z)=\left\{
\begin{array}{cl}
\left\{\cos\frac{\pi z}{2 \lambda},\;0,\; \sin\frac{\pi z}{2 \lambda}\right\},&
{\rm for}\; |z|< \lambda \\
\left\{\;0 \; ,\;\; 0\;\; , \;{\rm sgn}(z)\; \right\}, &{\rm for}\; 
|z| \ge \lambda 
\end{array}\right.
\end{equation}
admits an exact solution for the wave-function inside the domain wall
\cite{brataas} which has recently been used to calculate the torque 
that is due to a spin-polarized current \cite{waintal}. We use the
exact solution to determine the scattering properties of the domain
wall. Details are presented in Appendix \ref{sec:exact_trigo}.

By comparing with this exact analytic solution we checked the  
accuracy of the numerical method, as well as the absence of lattice 
effects for the parameters that we work with.

\section{Weak coupling}\label{sec:weak_coupling} 

The problem can be treated at the analytical level 
when the differential equations (\ref{eq:cde}) are only weakly
coupled. This is the case when the domain wall (in which the
spin-flip terms $f_x$ are non-zero) is very short, or when the
longitudinal energy of the electron is very high such that the transmitted
electrons spend only a short time inside the domain wall region. 
 
Then, we treat the system of coupled differential equations (\ref{eq:cde}) 
iteratively, considering the spin-flip terms on the right-hand-side as a
perturbation. Within this approach, the
solution to the homogeneous differential equation (\ref{eq:df1})
(in which the spin-flip terms induced by $f_x$ are set to zero) is
injected in the spin-flip term of the second differential 
equation (\ref{eq:df2}). This method was used in 
Ref.~[\onlinecite{wsj_moriond}], 
where a mechanism of channel blocking by a domain wall was proposed 
as a source of resistance in short ferromagnetic quantum wires. 
The starting point of this approach, which we present here for the
example of a linear domain wall as described by (\ref{eq:linear_wall}), 
is an incoming majority (spin-up) electron from the left 
$\phi_\uparrow^\mathrm{H}$, and $\phi_\downarrow^\mathrm{H}=0$. 
Outside the domain wall region, $\phi_\uparrow^\mathrm{H}$  reads
\begin{subequations}\label{eq:upoutside}
\begin{eqnarray}
\phi_{\uparrow}^\mathrm{H}(z)=&e^{ikz}+r_{\uparrow\uparrow}e^{-ikz}&\quad
\textrm{for}\quad z<-\lambda\\
\phi_{\uparrow}^\mathrm{H}(z)=& t_{\uparrow\uparrow}e^{ik'z}&\quad
\textrm{for}\quad z>\lambda \, ,
\end{eqnarray}
\end{subequations}
with the wave-numbers 
\begin{subequations}
\begin{eqnarray}
k&=&\sqrt{\frac{2m}{\hbar^2}\left(\epsilon+\frac{\Delta}{2}\right)}\\
k'&=&\sqrt{\frac{2m}{\hbar^2}\left(\epsilon-\frac{\Delta}{2}\right)}\, .
\end{eqnarray}
\end{subequations}
For $-\lambda<z<\lambda$, the homogeneous solution of Eq.\ (\ref{eq:df1}) is
\begin{eqnarray}
\label{eq:phiH}
\phi_{\uparrow}^\mathrm{H}(z) &=& 
\alpha \, \mathrm{Ai}\left[p^{2/3}\left(-\frac{2\epsilon}{\Delta} 
                + \frac{z}{\lambda}  \right)\right]\nonumber \\
&+& \beta \, \mathrm{Bi}\left[p^{2/3}\left(-\frac{2\epsilon}{\Delta} 
                + \frac{z}{\lambda}  \right)\right] 
\end{eqnarray}
with the usual Airy functions Ai and Bi, and the dimensionless 
parameter $p$ defined by
\begin{equation}
\label{p}
p=\left( \frac{m}{\hbar^2}\Delta \right)^{1/2} \lambda  
= \left( \frac{\Delta}{2E_{\textrm{F}}}\right)^{1/2} k_{\textrm{F}}\lambda \, .
\end{equation}
The coefficients $\alpha$ and $\beta$, as well as
$t_{\uparrow\uparrow}$ and $r_{\uparrow\uparrow}$, are obtained from  
the matching of the expressions (\ref{eq:upoutside}) and
(\ref{eq:phiH}) at $z=\pm \lambda$. For $\epsilon < \Delta/2$, 
we have imaginary $k'$. The transmission without spin-flip is zero
and $|r_{\uparrow\uparrow}|=1$. 
For $\epsilon > \Delta/2$, the transmission without spin-flip is finite. 

The first-order correction from the spin-conserving scattering is then 
obtained by injecting $\phi_\uparrow^\mathrm{H}$ as the inhomogeneous
term in the differential equation (\ref{eq:df2}) for $\phi_\downarrow$. 
Since we do not have incoming spin-down electrons, outside the domain
wall region we take the outgoing plane waves 
\begin{subequations}\label{eq:downoutside}
\begin{eqnarray}
\phi_{\downarrow}^{(1)}(z)=&r_{\uparrow\downarrow}e^{-ik'z}&\quad
\textrm{for}\quad z<-\lambda\\
\phi_{\downarrow}^{(1)}(z)=&t_{\uparrow\downarrow}e^{ikz}&\quad
\textrm{for}\quad z>\lambda \, .
\end{eqnarray}
\end{subequations}
For $-\lambda<z<\lambda$ the general solution can be written as
a linear combination of Airy functions plus a particular solution 
$\phi^\textrm{p}(z)$
\begin{eqnarray}
\label{eq:phi1}
\phi_{\downarrow}^{(1)}(z) &=& \alpha_1 \, 
\mathrm{Ai}\left[-p^{2/3}\left(\frac{2\epsilon}{\Delta}
+\frac{z}{\lambda}\right)\right] \nonumber \\
&+& \beta_1 \,
\mathrm{Bi}\left[-p^{2/3}\left(\frac{2\epsilon}{\Delta}
+\frac{z}{\lambda}\right)\right]
+ \phi^\textrm{p}(z) \, .
\end{eqnarray}
Spin-flip processes are now included in the description, and 
electrons undergoing a spin-flip can be transmitted even for energies 
$\epsilon<\Delta/2$. 
The parameters $\alpha_1$ and $\beta_1$ and the coefficients 
$t_{\uparrow\downarrow}$ and $r_{\uparrow\downarrow}$
are determined from the matching of 
(\ref{eq:downoutside}) and (\ref{eq:phi1}).

\subsection{Long wavelength limit}
\label{sec:longwl}

\begin{figure}
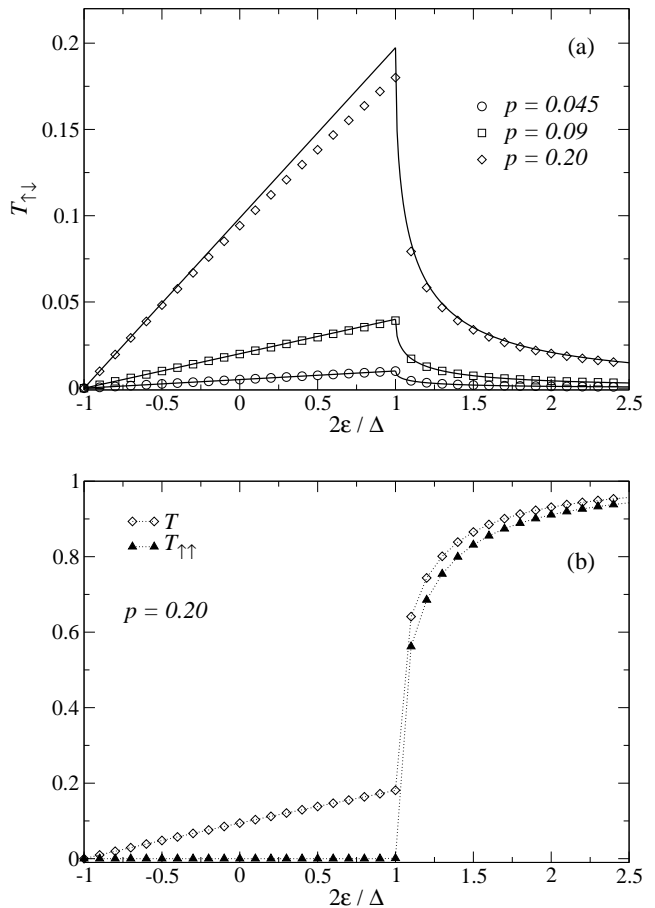

\begin{center}
\includegraphics[width=\columnwidth]{fig2a-GWJS.eps}\\
\includegraphics[width=\columnwidth]{fig2b-GWJS.eps}
\end{center}
\caption{\label{fig:perturbative}(a) Perturbative results for 
$T_{\uparrow\downarrow}$ from Eqs.\ (\ref{eq:tupdown_below}) and 
(\ref{eq:tupdown_above}) (solid lines) for three 
values of $p$, compared with the corresponding full numerical results 
(circles, squares and diamonds) for the linear domain wall. 
(b) The total transmission $T$ and 
$T_{\uparrow\uparrow}$ (diamonds and triangles, respectively) for $p=0.20$.}
\end{figure}
In the limit of a short wall, when the wavelength of the incoming
electron is much longer than the domain wall, the linear
approximation of the Airy functions allows us to write for $\epsilon
> \Delta/2$ 
\begin{subequations}
\begin{eqnarray}
T_{\uparrow \uparrow}&=& 
|t_{\uparrow \uparrow}|^2=\frac{4k k'}{(k + k')^2
+ 4(\lambda k k')^2}\\  
R_{\uparrow \uparrow}&=&|r_{\uparrow \uparrow}|^2=
\frac{(k-k')^2+4(\lambda k k')^2}{(k + k')^2
+4(\lambda k k')^2} \, .
\end{eqnarray}
\end{subequations} 
Obviously, for $\lambda\to 0$ we recover the well-known 
results for a step potential \cite{merzbacher}.

For $\epsilon < \Delta/2$, the transmission probability with spin-flip is
given by
\begin{equation}
\label{eq:tupdown_below}
T_{\uparrow\downarrow}=|t_{\uparrow\downarrow}|^2
=C^2 p^2\left( 1+\frac{2\epsilon}{\Delta} \right) \, . 
\end{equation}
For $\epsilon > \Delta/2$, 
\begin{equation}
\label{eq:tupdown_above}
T_{\uparrow\downarrow}=4C^2p^4\frac{4(x^2-1)+
p^2(x+1)(\frac{2}{3}(x-1)-1/p^2)^2}{\left((\sqrt{x+1}+\sqrt{x-1})^2+
4p^2(x^2-1)\right)^2} 
\end{equation}
is obtained, with $x=2\epsilon/\Delta$ and the prefactor $C$ defined by 
\begin{equation}
C=\frac{1}{\lambda}\int_{-\infty}^{\infty}\textrm{d}z\, f_x(z)\, .
\end{equation} 

In Fig.\ \ref{fig:perturbative} (a) we show $T_{\uparrow\downarrow}$ from 
Eqs.\ (\ref{eq:tupdown_below}) and (\ref{eq:tupdown_above}) as a
function of $2\epsilon/\Delta$, together with numerical calculations 
for three different values of $p$. We can see an excellent agreement
for the smallest values of $p$. When the value
of $p$ is increased, deviations appear first  
at energies close to $\Delta/2$. 
These features are consistent with the fact that the linear
approximation of the Airy functions is justified in the small 
$p$ limit, and becomes increasingly better for small energies.

The total transmission  
$T=T_{\uparrow\downarrow}+T_{\uparrow\uparrow}$
(Fig.~\ref{fig:perturbative} (b)) 
is dominated by the large transmission without spin-flip for 
$\epsilon>\Delta/2$. This feature justifies the ``channel blocking''
picture proposed for short domain walls in Ref.\ [\onlinecite{wsj_moriond}],
where the presence of the wall suppresses almost completely the
transmission at energies $\epsilon<\Delta/2$ ($T_{\uparrow\downarrow}$
is only a small correction for $\epsilon<\Delta/2$ and negligible for 
$\epsilon\gg \Delta/2$). Neglecting what happens for $\epsilon >
\Delta/2$, we have 
$\delta g/g=(-1+(Cp)^2)\Delta/2E_{\rm F}$.   
\begin{figure}
\begin{center}
\includegraphics[width=\columnwidth]{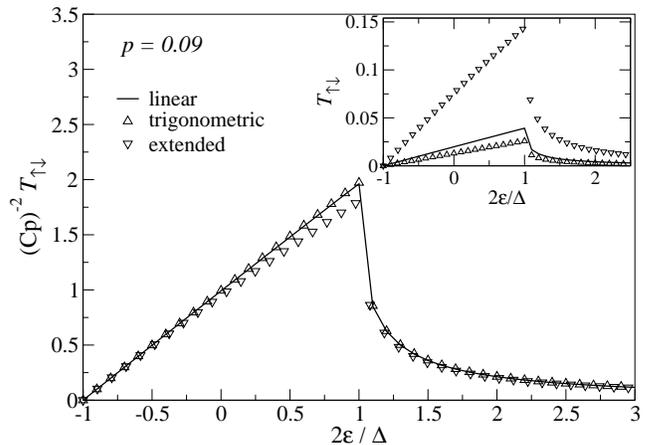}
\end{center}
\caption{\label{fig:universality}The transmission
$T_{\uparrow\downarrow}$ divided by the coupling strength 
$(Cp)^2$ for the linear, trigonometric and 
extended domain walls (solid line, up and down triangles, respectively).
The inset shows the results for $T_{\uparrow\downarrow}$ for the same value 
of $p=0.09$, before dividing by the corresponding value of $(Cp)^2$.}
\end{figure}

In the limit of short domain walls, we have found that the
dependence of the transmission coefficients on the shape of the wall
is only through the integral over $f_x$ which
enters in the prefactor $C$. For the linear, trigonometric and extended
domain walls, $C$ takes the values $\pi/2$, $4/\pi$ and $\pi$,
respectively. Such a scaling is shown in
Fig.\ \ref{fig:universality}, where the transmissions  
$T_{\uparrow\downarrow}$ divided by the coupling strength $(Cp)^2$ for 
the different domain wall shapes coincide for all energies, except
those close to $\Delta/2$. 
\begin{figure}
\begin{center}
\includegraphics[width=\columnwidth]{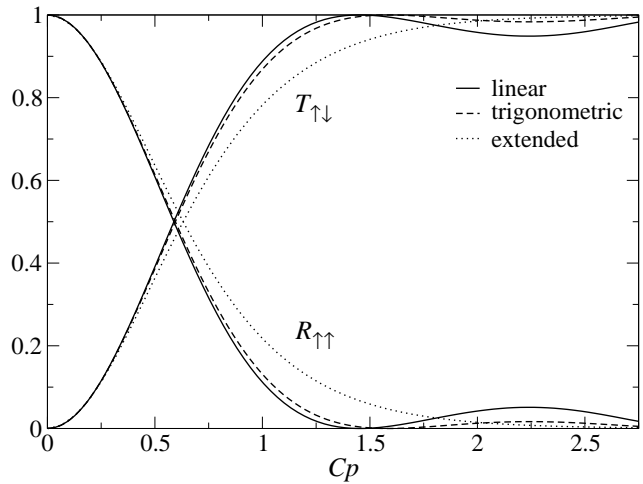}
\end{center}
\caption{\label{fig:oscillations}The coefficients $T_{\uparrow\downarrow}$ 
and $R_{\uparrow\uparrow}$ as a function of $Cp$ for the three
different domain wall shapes, at $\epsilon=0.95\Delta/2$. 
An oscillatory behavior for the linear and trigonometric walls is 
found as a consequence of the edges at the connection to the leads.}
\end{figure}

\subsection{Short wavelength limit}

The perturbative approach is not only applicable for short walls and
low energies (as in section \ref{sec:longwl}), but for general domain
wall parameters as well, provided that 
$\epsilon\gg (\Delta\lambda/\hbar)^2/m$.
That is, when the time that the electron 
spends inside the domain wall is
much shorter than the spin precession period, and therefore the
spin-flips are very unlikely. In this limit, the WKB approximation of
the scattering wave-functions for the linear domain wall model 
(Eq.~(\ref{eq:linear_wall})) yields the reflection and transmission 
coefficients
\begin{eqnarray}
R_{\uparrow\uparrow}&=& \left(\frac{\Delta}{4\epsilon}\right)^2
\frac{\sin^2(2k\lambda)}{(2k\lambda)^2}\\
R_{\uparrow\downarrow}&=& \left(\frac{C\Delta}{8\epsilon}\right)^2
\sin^2(2k\lambda)\\
T_{\uparrow\downarrow}&=& \left(\frac{C\Delta}{8\epsilon}\right)^2
(2k\lambda)^2\\
T_{\uparrow\uparrow}&=& 1-R_{\uparrow\uparrow}-R_{\uparrow\downarrow}-
T_{\uparrow\downarrow}
 \, .
\end{eqnarray}
Thus, for energies $\epsilon\gg\Delta$ all scattering
coefficients, except the transmission without spin-flip, are very
small. Therefore, in first approximation we can neglect the effect of
the domain wall for electrons with high longitudinal energies. The
conductance associated with the domain wall is then determined by the
low-energy electrons \cite{wsj_moriond}. The algebraic decay in 
$\Delta/\epsilon$ is less pronounced than the exponential suppression
obtained by Cabrera and Falicov \cite{cabrera}. Such a difference
arises from the sharp edges at $z=\pm\lambda$ in the linear domain
wall model we used for this calculation.

\section{Strong coupling}\label{sec:strong_coupling}

\begin{figure}
\begin{center}
\includegraphics[width=\columnwidth]{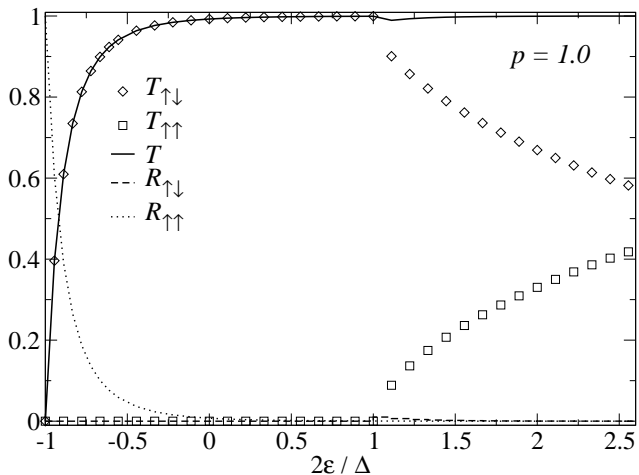}
\end{center}
\caption{\label{fig:intermediate}Transmission and reflection
coefficients in the intermediate regime ($p=1$) for the 
extended domain wall geometry (\ref{eq:extended_wall}). 
A large transmission with spin-flip (diamonds) is found for 
energies $\epsilon<\Delta/2$ in this regime, where the transport is adiabatic.}
\end{figure}
If we are interested in energies
$\epsilon\simeq\Delta/2$ and not necessarily short walls, the previous
picture has to be modified. The linear approximation and the
perturbative treatment (involving only one spin-flip) in the wall
region are no longer justified. Beyond the perturbative regime, the detailed
shape of the domain wall might become relevant.

In Fig.\ \ref{fig:oscillations}, we present $T_{\uparrow\downarrow}$
and $R_{\uparrow\uparrow}$ as a function of the coupling strength for
different domain wall shapes and an energy of the order of $\Delta/2$.
As discussed in the previous section, it is for these energies
that a dependence on the detailed shape of the domain wall 
appears first when departing from the weak coupling limit ($Cp\ll 1$). 
We can see from Fig. \ref{fig:oscillations} that the transmissions 
(reflections) for the different domain wall shapes coincide for small 
values of $Cp$. Even for stronger couplings, the different models do 
not show very important differences in their behaviors. The only
apparent difference are oscillations of the transmission coefficients 
as a function of $p$,
which occur at intermediate $p$ for linear and trigonometric domain 
walls. The origin of these oscillations is due to the
edges of the domain wall region leading to Fabry-Perot like
interferences. For a smooth domain wall structure such as 
the extended domain wall, the oscillations are absent. 
On the other hand, and as expected, $T_{\uparrow\downarrow}\to
1$ for all shapes in the limit of large $p$. 

It is the limit of infinite domain wall length where the spin follows
adiabatically the orientation of the local magnetization \cite{stern},
corresponding to a rotation from spin up to spin down in the
external basis of fixed spin orientations. Electrons are transmitted
with probability one through the wall, therefore
$T_{\uparrow\downarrow}=1$ and 
$T_{\uparrow\uparrow}=R_{\uparrow\uparrow}=R_{\uparrow\downarrow}=0$.
In this limit the detailed shape of a domain wall, having slow
spatial spin rotation, is irrelevant. The adjustment of the spin to the 
direction of the local magnetization requires an infinite number of
spin-flips (in the fixed basis), and obviously cannot be described by
taking into account a small number of spin-flips as in the
perturbative approach used for short domain walls. The condition for
the local adjustment is that the Larmor precession of the spin around
the local magnetization is much faster than the rotation of the local
magnetization viewed by the traveling electron \cite{stern}. 
This condition of adiabaticity translates into 
$\Delta\gg(h/\lambda)\sqrt{\epsilon/m}$.

\begin{figure}
\begin{center}
\includegraphics[width=\columnwidth]{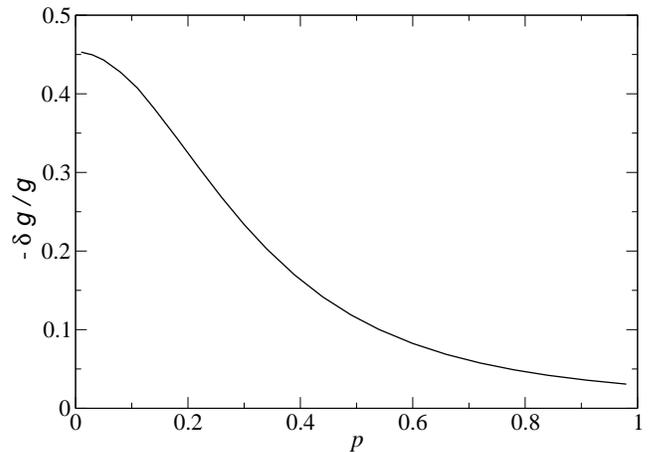}
\end{center}
\caption{\label{fig:integrated} The relative change in conductance 
$\delta g/g$ caused by the presence of a domain wall (extended shape), 
for $E_{\rm F}=2\Delta$.}
\end{figure}
We then see that the adiabatic condition strongly depends on the
longitudinal kinetic energy of the electrons. In a quantum wire, at a
finite value of $\lambda$, electrons with low longitudinal velocity
are essentially adiabatic, while the channels with low transverse
quantum numbers can be highly non-adiabatic. 
In calculating the conductance of a ferromagnetic quantum wire, we
have to take into account the co-existence of adiabatic (low
longitudinal energy) and non-adiabatic (high longitudinal energy) 
electrons. It then seems important to work out the
crossover between the short wall and adiabatic limits, for different
shapes of the domain wall.

For an intermediate value of $p$ in the case of an extended domain
wall, Fig.\ \ref{fig:intermediate} shows that the behavior for
$\epsilon<\Delta/2$ is radically different from the weak coupling
case of section \ref{sec:weak_coupling}. The weak coupling result of
Eq.\ (\ref{eq:tupdown_below}) is only valid at extremely low energies,
and the $T_{\uparrow\downarrow}$ approaches one (adiabatic behavior)
for longitudinal energies considerably lower than the step height.
Above $\Delta/2$, $T_{\uparrow\downarrow}$ decreases monotonously with
energy, returning to the weak coupling regime in the limit of large
$\epsilon$. At the same time, $T_{\uparrow\uparrow}$ increases towards
one and $T$ remains very close to perfect transmission for all
energies.

Thus, almost all of the electrons with energy $\epsilon > \Delta/2$ are 
transmitted. However, while the spin of the transmitted electrons is
changed by the domain wall for low $\epsilon$ and high $p$, the spin of
electrons having high $\epsilon$ in domain walls of low $p$ remains
unaffected by the wall (see also Fig.\ \ref{fig:perturbative} (a)). 
Therefore, in calculating the effect of the
domain wall on the quantum conductance the modes with longitudinal
energies in the interval $(-\Delta/2,\Delta/2)$ are most relevant 
\cite{wsj_moriond}. 

The conductance for the ideal ballistic case given in Eq.\
(\ref{eq:landauer}) is obtained by summing
over all conductance channels. Fig.\ \ref{fig:integrated} shows
an example of the resulting behavior for the 
difference in conductance between the cases without and with domain
wall (normalized to the conductance without domain wall), as a
function of the domain wall parameter $p$. We can see that the
channel blocking effect due to the presence of the domain wall is
rapidly suppressed upon increase of the coupling. Similar results have
recently been obtained using a different numerical approach \cite{falloon}.

\section{Summary and Conclusions}

The effect of a single domain wall on the electronic transport in a
ferromagnetic nanowire has been studied systematically in various
parameter regimes. The domain wall leads to a coupling of spin up and
spin down electrons in the conduction channels, which is proportional
to the exchange energy of the conduction electrons and the length of
the wall. 

For an abrupt domain wall the step in the effective potential felt by the
conduction electrons blocks the transmission of channels with low
longitudinal energy. In the weak coupling limit, a perturbative
approach is possible, leading to the lowest order correction to
perfect channel blocking. In this case, the detailed shape of the
domain wall is not relevant, and the transmission coefficients scale
with the coupling strength.
For a very long domain wall the spin of the electrons follows adiabatically
the local effective magnetization and the conductance is unaffected by
the domain wall, independently of its shape.

The intermediate coupling regime is most relevant for the domain walls
that can be investigated experimentally. We have shown that the degree
of adiabaticity of electrons at the Fermi energy strongly depends on
their longitudinal kinetic energy. While the spin of electrons with low
longitudinal energy essentially behaves adiabatically, the spin of
electrons with high longitudinal energy is practically unaffected by
the domain wall. The crossover between these two behaviors, as a
function of the longitudinal energy of the electrons, has to be taken
into account in calculating the conductance of the quantum wire.

Our analysis has been based on coherent scattering at the domain wall,
which is connected to perfect leads. However, in realistic situations,
the domain wall is not connected to scattering-free regions. 
The imperfections and impurities at both sides of the wall give rise
to elastic scattering, which may be different for the two spin
directions of the electrons. Though these coherent effects can in 
principle be taken into account in a scattering approach, such a coherent
picture is not sufficient for wires which are longer than the 
phase coherence length. Since this is the case in typical experiments, 
we need in addition to take into account inelastic processes (like
electron-phonon or spin-magnon scattering). The length of the leads
over which the spin of the electrons is conserved can then be
described phenomenologically by classical spin-dependent resistors 
\cite{imry}. In this situation, the important part of the electrons 
which do not undergo spin-flip processes leads to an increase of the 
resistance due to the GMR mechanism. This picture is likely to be 
representative of the experimental situations [\onlinecite{ebels,dumpich}].
More experimental and theoretical work concerning the various
relaxation rates will be necessary to establish a complete
quantitative understanding of the phenomenon. 

\begin{acknowledgments}
We thank P.\ Falloon, H.\ Pastawski and X.\ Waintal for very useful 
discussions. In addition, we are grateful to H.\ Pastawski for crucial 
help in the implementation of the numerical method, and to X.\ Waintal 
for drawing our attention to the exact solution for the spin spiral of
Ref.~\onlinecite{brataas}. This work received financial support from
the European Union within the RTN program (Contract No.\ HPRN-CT-2000-00144). 
V.G.\ thanks the French Minist\`ere d\'el\'egu\'e \`a la recherche et
aux nouvelles technologies and the Center for Functional
Nanostructures of the Deutsche Forschungsgemeinschaft (project B2.10) for
support.
\end{acknowledgments}

\appendix

\section{Exact solution for the trigonometric domain wall}
\label{sec:exact_trigo}

A particularly instructive case is that of a trigonometric domain wall 
(Eq.~(\ref{eq:trigonometric_wall})) since an exact solution for the
wave function inside the domain wall region can be obtained \cite{brataas}.
Here we extend this approach to a scattering situation by matching the
inner solutions with plane waves, which allows us to calculate the
transmission and reflection amplitudes.

In addition to the external spin basis 
$\{|z,\uparrow\rangle , \, |z,\downarrow\rangle\}$, 
it is useful to introduce a local spin basis $\{|z,\uparrow^{\rm L}\rangle , 
|z,\downarrow^{\rm L}\rangle\}$, 
which corresponds to spin orientations parallel and anti-parallel 
to the (rotating) local magnetization direction $\vec{f}(z)$, leading to
\begin{equation}
\left(\begin{array}{c}|z,\uparrow^{\rm L}\rangle\\ 
|z,\downarrow^{\rm L}\rangle  \end{array}\right)
=R(z)\left(\begin{array}{c}
|z,\uparrow\rangle \\ 
|z,\downarrow\rangle \end{array}\right)
\end{equation}
where the spin rotation matrix is given by 
\begin{equation}
R(z)=\left(\begin{array}{cc}
\cos(az+\pi/4)&\sin(az+\pi/4)\\
-\sin(az+\pi/4)&\cos(az+\pi/4)
\end{array}\right)
\end{equation}
with $a=\pi/4\lambda$. For $z=-\lambda$, $R$ is simply the identity
matrix (the local rotating basis coincides with the fixed one), and
putting $z=\lambda$ corresponds to exchanging the spin directions
between the local and fixed bases.

Inserting the spinor 
\begin{equation}
|\psi(z)\rangle=\phi^{\rm L}_\uparrow(z)|z,\uparrow^{\rm L}\rangle
+\phi^{\rm L}_\downarrow(z)|z,\downarrow^{\rm L}\rangle
\end{equation} 
into the Schr\"odinger equation corresponding to the Hamiltonian 
(\ref{eq:H}), we obtain
\begin{subequations}
\label{eq:cde_in_local_basis}
\begin{equation}
\label{eq:df1_local}
\left[\frac{d^2}{dz^2}-a^2\right]\phi^{\rm L}_\uparrow+
\frac{2m}{\hbar^2}\left(\epsilon-\frac{\Delta}{2} \right)
\phi^{\rm L}_\uparrow =2a\frac{d}{dz} \phi^{\rm L}_\downarrow
\end{equation}
\begin{equation}
\label{eq:df2_local}
\left[\frac{d^2}{dz^2}-a^2\right]\phi^{\rm L}_\downarrow+
\frac{2m}{\hbar^2}\left(\epsilon+\frac{\Delta}{2} \right)
\phi^{\rm L}_\downarrow =-2a\frac{d}{dz} \phi^{\rm L}_\uparrow
\end{equation}
\end{subequations} 
which is, in fact, Eq.\ (\ref{eq:cde}) expressed in the local spin
basis, for the case of a trigonometric domain wall.
This system of coupled differential equations can be reduced to a 
$2\times 2$ eigenvalue problem with the ansatz 
\begin{equation}
\left(\begin{array}{c}\phi_\uparrow^{\rm L}(z)\\ 
\phi_\downarrow^{\rm L}(z)  \end{array}\right)
=\exp(i\tilde{k}z)\left(\begin{array}{c}
C_\uparrow \\ 
C_\downarrow \end{array}\right)\, ,
\end{equation}
such that the solutions $(C_\uparrow,C_\downarrow)$ and $2m\epsilon/\hbar^2$
are the eigenvectors and eigenvalues, respectively, of the matrix
\begin{equation}
M=\left(\begin{array}{cc}\tilde{k}^2+a^2+m\Delta/\hbar^2& 2i\tilde{k}a\\ 
-2i\tilde{k}a& \tilde{k}^2+a^2-m\Delta/\hbar^2  \end{array}\right)\, .
\end{equation}
The secular equation of $M$ leads to the dispersion relations
\begin{equation}
\epsilon_{1,2}=\frac{\hbar^2}{2m}\left(\tilde{k}^2+a^2\pm\sqrt{4a^2\tilde{k}^2+\frac{2m}{\hbar^2}\left(\frac{\Delta}{2}\right)^2}\right)
\end{equation}
that is, the eigenenergies of the infinite spin spiral
\cite{brataas}. For a closed spiral, the periodic boundary conditions
would lead \cite{stern} to quantized values of $\tilde{k}$. However, we are
interested in a scattering problem, where the region in which the
magnetization turns is connected to homogeneous ferromagnetic leads. 
We therefore express the general solution 
$(\phi_\uparrow^{\rm L},\phi_\downarrow^{\rm L})$ for a given energy
as a linear combination of the four corresponding eigenstates of the
spiral, and use the matching conditions between the domain wall region
and the perfect leads at $z=\pm \lambda$. Taking into account the
rotation of the local basis, and using the expressions given in Eqs.\
(\ref{eq:upoutside}) and (\ref{eq:downoutside}) for the wave-function
outside the wall region, we get
\begin{eqnarray}
e^{-ik\lambda}+r_{\uparrow\uparrow}e^{ik\lambda}
&=&\phi_\uparrow^{\rm L}(-\lambda)   \nonumber    \\
t_{\uparrow\uparrow}e^{ik'\lambda}
&=&-\phi_\downarrow^{\rm L}(\lambda) \nonumber    \\
r_{\uparrow\downarrow}e^{ik'\lambda}
&=&\phi_\downarrow^{\rm L}(-\lambda) \nonumber    \\
t_{\uparrow\downarrow}e^{ik\lambda}
&=&\phi_\uparrow^{\rm L}(\lambda)    \nonumber    \\
ik\left(e^{-ik\lambda}-r_{\uparrow\uparrow}e^{ik\lambda}\right)
&=&\frac{\rm d}{{\rm d}z}\phi_\uparrow^{\rm L}(-\lambda)
-a\phi_\downarrow^{\rm L}(-\lambda)  \nonumber    \\
ik' t_{\uparrow\uparrow}e^{ik'\lambda}
&=&-\frac{\rm d}{{\rm d}z}\phi_\downarrow^{\rm L}(\lambda)
-a\phi_\uparrow^{\rm L}(\lambda)      \nonumber   \\
-ik' r_{\uparrow\downarrow}e^{ik'\lambda}
&=&\frac{\rm d}{{\rm d}z}\phi_\downarrow^{\rm L}(-\lambda)
+a\phi_\uparrow^{\rm L}(-\lambda)     \nonumber   \\
ik t_{\uparrow\downarrow}e^{ik\lambda}
&=&\frac{\rm d}{{\rm d}z}\phi_\uparrow^{\rm
  L}(\lambda)
-a\phi_\downarrow^{\rm L}(\lambda)\, . \nonumber
\end{eqnarray}
These conditions allow us to extract the amplitudes
$t_{\uparrow\uparrow}$, $r_{\uparrow\uparrow}$,
$t_{\uparrow\downarrow}$ and $r_{\uparrow\downarrow}$, as well as the
precise form of the wave-function inside the domain wall. The
resulting transmission coefficients are presented and discussed in
Sec.\ \ref{sec:strong_coupling}.

\end{document}